# Intramolecular coupling of terminal alkynes by atom manipulation

Florian Albrecht*[a]#, Dulce Rey[b]#, Shadi Fatayer[a], Fabian Schulz[a], Dolores Pérez[b], Diego Peña*[b] and Leo Gross*[a]

In memory of Kilian Muñiz

[a]   Dr. F. Albrecht[#], Dr. S. Fatayer, Dr. F. Schulz, Dr. L. Gross
      IBM Research – Zurich
      CH-8803 Rüschlikon (Switzerland)
      E-mail: FAL@zurich.ibm.com, LGR@zurich.ibm.com

[b]   D. Rey[#], Dr. D. Pérez, Dr. D. Peña
      Centro de Investigación en Química Bioló xica e Materiais Moleculares (CiQUS) and
      Departamento de Química Orgánica, Universidade de Santiago de Compostela
      15782 Santiago de Compostela (Spain)
      E-mail: diego.pena@usc.es

[#]   These authors contributed equally to this work.

**Abstract:**

Glaser-like coupling of terminal alkynes by thermal activation is extensively used in on-surface chemistry. Here we demonstrate an intramolecular version of this reaction performed by atom manipulation. We used voltage pulses from the tip to trigger a Glaser-like coupling between terminal alkyne carbons within a custom synthesized precursor molecule adsorbed on bilayer NaCl on Cu(111). Different conformations of the precursor molecule and the product were characterized by molecular structure elucidation with atomic force microscopy and orbital density mapping with scanning tunneling microscopy, accompanied by density functional theory calculations. We revealed partially dehydrogenated intermediates providing insight into the reaction pathway.

On-surface synthesis has made great progress since the demonstration of covalent coupling by thermal activation pioneered by Grill *et al.* in 2007 [1]. Milestones include on-surface created polymers [2], graphene nanoribbons [3], a plethora of novel molecules and a wealth of one- and two-dimensional nanostructures with intriguing properties [4, 5]. By contrast, the on-surface covalent coupling by atom manipulation is still in its infancy, and only a few successful examples have been reported so far [6, 7].

On-surface chemistry by thermal activation, on the one hand, and by atomic manipulation, on the other hand, are complementary approaches with different advantages and challenges. In general, large defect-free structures with identical repeating units can be produced by thermal activation. With atomic manipulation, every single reaction step can be induced and studied separately [8], defects can be deliberately introduced, units do not need to repeat, and thus more complex custom designed structures far from equilibrium could be fabricated in the future. In addition to its flexibility, atom manipulation has the advantage that often reactions do not require the catalytic activity of the metallic substrate and can be performed on insulating films. On ultrathin insulating films, e.g. few layer NaCl or monolayer Xe, also reactive intermediates can be stabilized [7, 9] and structures are electronically decoupled, which is advantageous for their electronic characterization [10] and possible molecular electronics applications.

The oxidative coupling of terminal alkynes promoted by cuprous salts is a well-known and classic reaction in organic synthesis, frequently named as the Glaser-Hay coupling (Scheme 1a) [11]. Among the reactions successfully introduced in the field of on-surface synthesis, this coupling reaction occupies a prominent position. It was employed for building polymers on metallic surfaces [12, 13, 14, 15, 16, 17, 18] and recently was also observed on the surface of a bulk insulator by thermal activation [19]. However, this reaction has never been demonstrated by atom manipulation up to now. Here we report the first example of on-surface Glaser coupling by atom manipulation on single molecules.

Intermolecular coupling reactions are challenging to initiate by atom manipulation, in part because of the difficulty of aligning the precursor molecules. Performing such reactions intramolecularly, facilitates inducing the reaction, because the reacting functional groups can be pre-aligned by design. Intramolecular reaction-experiments can tell us if such coupling reactions as the Glaser reaction are in principle possible by atom manipulation and could reveal details about the reaction pathway and its intermediates. With this idea in mind we envisaged the synthesis of compound **1** as an ideal molecular model to explore this reaction by atom manipulation (Scheme 1b). The expected proximity between the terminal alkynes would facilitate the C-C bond formation (in red) to obtain the 1,8-anthrylene-ethynylene cyclic dimer **2**.

Although compound **2** was not previously reported, the synthesis of an alkyl substituted derivate had been achieved by Toyota and co-workers in 2007 by means of a double Sonogashira reaction in solution [20]. For the synthesis of compound **1** we modified a procedure developed by this group, starting with commercially available 1,8-dibromoanthracene **3** (Scheme 1c) [21]. Double Sonogashira coupling between compound **3** and trimethylsilylacetylene (TMS = trimethylsilyl), followed by the fluoride-induced deprotection of the alkynes and the selective introduction of one triisopropylsilyl (TIPS) group led to the isolation of diyne **4**. Then, an intermolecular Glaser-like homocoupling followed by fluoride-induced cleavage of the TIPS afforded compound **1** as a yellow solid.

The precursor molecules **1** were deposited via sublimation from a Si wafer onto a cold ($T \approx 10$ K) Cu(111) surface partially covered with (100) oriented bilayer NaCl islands. The experiments were carried out in a combined scanning tunneling microscope/atomic force microscope (STM/AFM) system equipped with a qPlus force sensor [22] operating at $T = 5$ K in frequency-modulation mode [23]. We used CO tip functionalization to improve the resolution for AFM [24] and STM [25]. AFM images were acquired at constant height, with the tip-height offset $\Delta z$ applied to the tip-sample distance with respect to the STM setpoint above the bare NaCl surface.

After deposition of **1** on the substrate at $T = 10$ K, we found three different conformations of the molecules on NaCl(2ML)/Cu(111), as shown in the CO-tip AFM images in Fig. 1b)-d). The majority (55 %) of molecules are found in the conformation shown in Fig. 1b), that we assign to *trans*-**1**. In this conformation the molecule is prochiral and both corresponding enantiomeric adsorbates are observed with about equal occurrences. The characteristic bright contrast that relates to the triple bonds [7, 26, 27] is observed in the AFM images. The similar brightness of the entire molecule and the symmetric AFM contrast indicate a planar adsorption geometry [28]. A slightly smaller set of molecules (about 45%) is found with the contrasts shown in Fig. 1c) and d), that we assign to molecules adsorbed in the *cis* conformations *(P)-cis*-**1** and *(M)-cis*-**1**. The asymmetric contrast, i.e., the different brightness observed for the two anthracene moieties, indicates a non-planar adsorption, explained by steric hindrance between the two terminal alkyne groups. This steric hindrance forces one terminal alkyne group being closer to the substrate (down) and the other one being further away (up), resulting in significantly brighter contrast above the latter. Thus, the *cis* conformation shows helicity and is chiral. Furthermore, the two anthracene moieties are not parallel to each other, probably also related to the steric hindrance between the end groups. These observations suggest that a carbon-carbon bond between the alkyne end groups had not formed and does not form spontaneously at these cryogenic temperatures. As expected, we observed both enantiomers *(P)-cis*-**1** and *(M)-cis*-**1** with about equal occurrences.

It was possible to switch the helicity of compound *cis*-**1** [29, 30], i.e., between *(P)-cis*-**1** and *(M)-cis*-**1** in both directions. Such switching occurred frequently during AFM imaging when the tip height was decreased with respect to the tip height in Fig. 1c), d) (see Supporting Information). We never observed a conformational change between the *trans*-**1** and *cis*-**1** conformations on the surface.

To study if Glaser-like coupling reactions can be induced by atom manipulation we focused on the *cis*-**1** conformers, in which the end groups are aligned and positioned to facilitate a possible coupling of the terminal alkynes. To induce the reaction, the tip was positioned above such a molecule on bilayer NaCl on Cu(111) and retracted by 9 - 11 Å from a tunneling setpoint of $V = 1$ V and $I = 1$ pA. Then the sample voltage was ramped up to sample voltages $V > 4.5$ V ($V = 5.6$ V in the case shown in Fig. 1) resulting in currents of up to 400 pA. The molecule typically changed its location slightly during the voltage pulse and thus the product had to be identified from a subsequent image. Successive AFM imaging, after such manipulation of a *cis*-**1** precursor, revealed the molecule shown in Fig. 1e), f) that we assign to the Glaser coupled product **2**.

We investigated the electronic structure of **2** by obtaining STM images at the negative ion resonances, which can be related to orbital density maps of the lowest unoccupied molecular orbitals (LUMOs) of the molecule [10]. The positive ion resonance, related to the highest occupied molecular orbital, could not be experimentally accessed. In differential conductance (d$I$/d$V$) spectroscopy, we observed peaks centered at +1.5 V, +2.0 V and a third one with onset at +2.3 V, that we assign to resonant tunneling into the LUMO, LUMO+1 and LUMO+2, respectively (see Supporting Information). We obtained STM constant-current images at these voltages using a CO terminated tip, i.e., a *p*-wave tip [25], see Fig. 2 first row, and a Cu tip, i.e., an *s*-wave tip, see Fig. 2 second row.

Additionally, we carried out density functional theory (DFT) calculations shown in Fig. 2, third row. The orbitals of **2** can be rationalized as those of two coupled diethynylanthracene molecules, with the LUMO and LUMO+1 of **2** displaying a bonding and an antibonding combination [31] of two diethynylanthracene LUMOs, respectively and the LUMO+2 of **2** displaying a bonding combination of two diethynylanthracene LUMO+1 orbitals (see Supporting Information).

To compare with the images obtained with the Cu (*s*-wave) tip, we folded the calculated wavefunction with a spherical function to account for the *s*-wave character and finite size of the tip [32], see Fig. 2 fourth row. The agreement of the calculated wavefunctions with the measured orbital densities (corresponding to wavefunctions squared) is excellent, confirming the creation of **2** by atom manipulation.

A theoretical study of the reaction mechanism of the thermally induced Glaser-like coupling on Ag(111) indicated that the carbon-carbon bond is formed first and then dehydrogenation occurs, with the latter being the limiting step of the reaction [33]. Such intermediates, after carbon-carbon bond formation but before dehydrogenation have been observed in on-surface cyclodehydrogenation [34] but not in Glaser-like coupling reactions. After performing voltage pulses above *cis*-**1** we once observed reaction intermediates **5** and **6**, both with an additional hydrogen atom (with respect to **2**) still attached to the molecule, see Figure 3b) and 3c), respectively. This observation suggests that carbon-carbon bond formation occurred before dehydrogenation - at least of the second hydrogen atom (in red in Figure 3a). It also suggests that the second hydrogen, which still needs to be dissociated form **2**, is relatively mobile within the molecule. Supporting the latter point, i.e., the possible conversion between **5** and **6**, we observed that the position of the additional hydrogen could be altered by atom manipulation. A voltage pulse ($V$ = 4 V) resulted in a change of the position of the additional hydrogen from the anthracene moiety in **5**, see Fig. 3b), to the acetylene bridge in **6**, see Fig. 3c). When we applied a voltage pulse of 4.5 V above **6** in Fig. 3c), it resulted in dehydrogenation and completed the synthesis of **2**, see Fig. 3d).

Our results demonstrate that Glaser-like coupling reactions can be induced by atom manipulation and revealed intermediates indicating that the second dehydrogenation step occurred after carbon-carbon bond formation. This work shows that the proper design of the molecular precursor can be crucial to facilitate on-surface reactions by atom manipulation, providing unprecedented insights into the reaction mechanisms in single molecules.

**Acknowledgements:**

We thank Katharina Kaiser, Enrique Guitián and Rolf Allenspach for discussions. We thank the European Union (Project SPRING, contract no. 863098), the ERC grant AMSEL (682144), the Spanish Agencia Estatal de Investigación (MAT2016-78293-C6-3-R and CTQ2016-78157-R), Xunta de Galicia (Centro de Investigación de Galicia accreditation 2019–2022, ED431G 2019/03) and the European Regional Development Fund-ERDF for financial support.

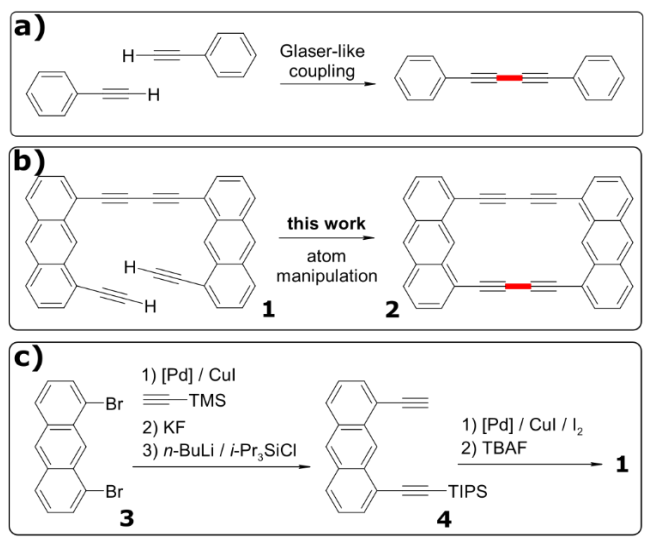

**Scheme 1.** a) Intermolecular Glaser-like coupling. b) Intramolecular Glaser-like coupling. c) Synthesis of compound **1**.

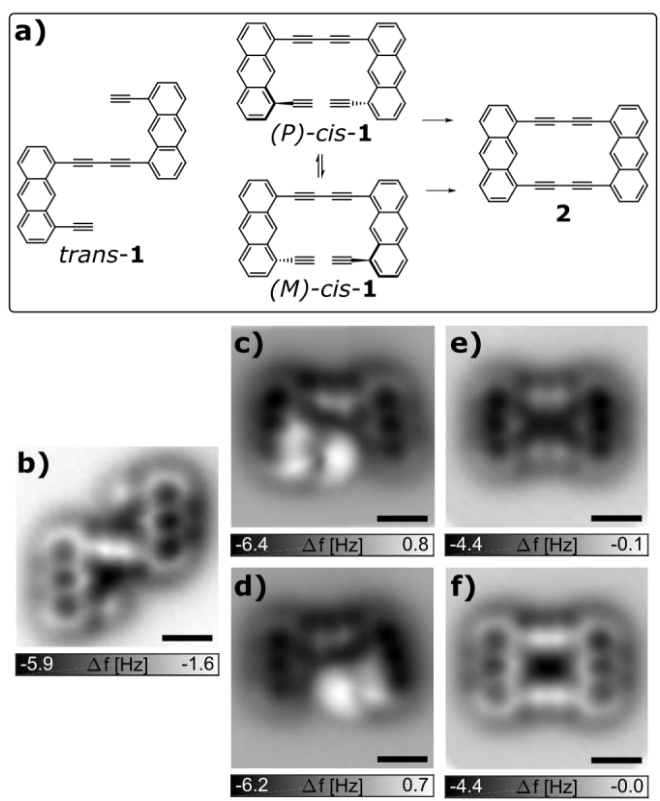

*Figure 1.* **Precursors and product.** a) Schemes of the precursor molecule **1** in the *trans* and *cis* conformations, and the product **2**. b) Constant-height AFM image with a CO functionalized tip on a precursor in the *trans* conformation on bilayer NaCl on Cu(111), tip-height offset $\Delta z = 0.85$ Å, i. e., the tip-sample spacing was increased by 0.85 Å from a STM setpoint of $I = 1$ pA and $V = 0.1$ V. Panels c) and d) show two enantiomeric conformations of *cis*-**1**, before and after its helicity was changed from *(P)* to *(M)* by atom manipulation, $\Delta z = 1.0$ Å in c) and $\Delta z = 1.1$ Å in d). Panels e) and f) show the product after a tip-induced Glaser coupling reaction, imaged at tip-height offsets of $\Delta z = 1.7$ Å in e) and $\Delta z = 1.5$ Å in f). The data shown in e) and f) were recorded with a different tip than b)-d). All scale bars are 5 Å.

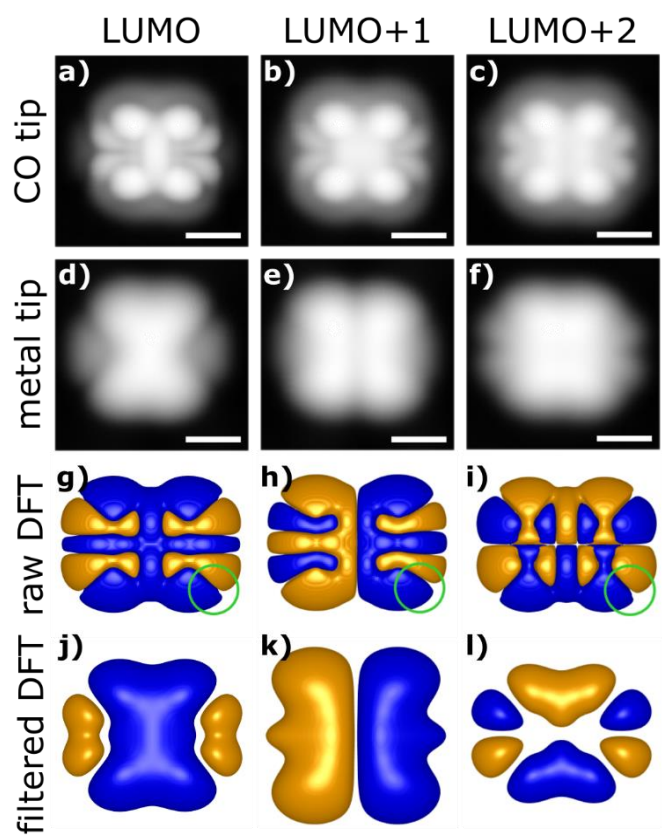

*Figure 2.* **Orbital densities of reaction product**. STM orbital-density maps and calculations of the LUMO, LUMO+1 and LUMO+2 densities of **2**. STM constant-current images recorded at a current of 1 pA with a CO functionalized tip (first row, at bias voltages of 1.5 V, 2.0 V and 2.5 V) and with a Cu tip (second row, 1.5 V, 2.1 V and 2.5 V) on **2** adsorbed on NaCl(2ML)/Cu(111). Black to white contrast corresponds to tip-height ranges of 3.5 Å for panels a) and f) and of 4.0 Å for panels b) to e). All scale bars correspond to 5 Å. Row three shows the corresponding iso-surfaces of calculated orbital wavefunctions. The bottom row shows iso-surfaces of DFT calculated orbitals after a 3D Gaussian filter was applied to mimic the spatial extension of the metallic tip. The green circles in the third row indicate the full width half maximum of this 3D Gaussian filter.

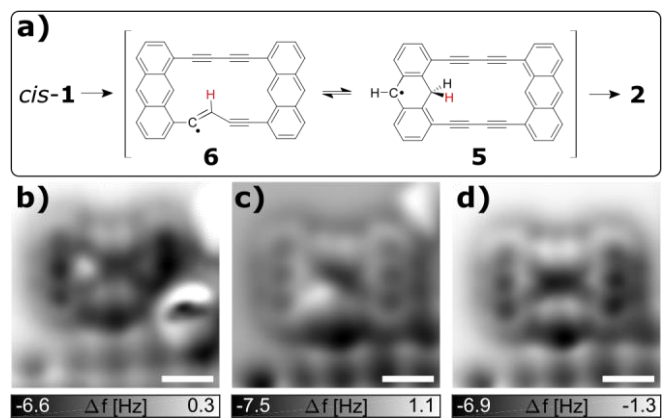

**Figure 3. Reaction intermediates.** a) Reaction scheme with observed intermediates. b)-d) Constant-height AFM with CO tip. The molecule was anchored next to a third layer patch of NaCl (bottom of images). Between recording b), assigned as **5**, and c), assigned as **6**, a voltage pulse of 4 V was applied. Between recording c) and d), assigned as **2**, a voltage pulse of 4.5 V was applied. Constant-height AFM with CO tip and tip-height offsets Δz = 1.4 Å in a), 1.0 Å in b) and 1.1 Å in c) with respect to the STM setpoint of 1 pA and 0.1 V. All scale bars correspond to 5 Å.

# Supporting Information

# Intramolecular coupling of terminal alkynes by atom manipulation

Florian Albrecht, Dulce Rey, Shadi Fatayer, Fabian Schulz, Dolores Pérez, Diego Peña, Leo Gross

**Synthesis of precursor molecules 1:**

*General methods for the solution synthesis*

All reactions were carried out under argon using oven-dried glassware. *n*-BuLi was used in solution in hexane (2.4 M). TBAF was used in solution in THF (1 M). Et$_3$N, *i*-Pr$_2$NH and *i*-Pr$_3$SiCl were dried by distillation over CaH$_2$. Commercial reagents were purchased from ABCR GmbH, Sigma-Aldrich or TCI Chemicals and were used without further purification. THF and CH$_2$Cl$_2$ were purified by a MBraun SPS-800 Solvent Purification System. TLC was performed on Merck silica gel 60 F$_{254}$ and chromatograms were visualized with UV light (254 and 365 nm). Flash column chromatography was performed on Merck silica gel 60 (ASTM 230-400 mesh). $^1$H and $^{13}$C NMR spectra were recorded at 300 and 75 MHz, respectively (Varian Mercury 300). APCI spectra were determined on a Bruker Microtof instrument. Deuterated solvents were purchased from Acros Organics.

*Preparation of 1,8-bis((trimethylsilyl)ethynyl)anthracene (7)*

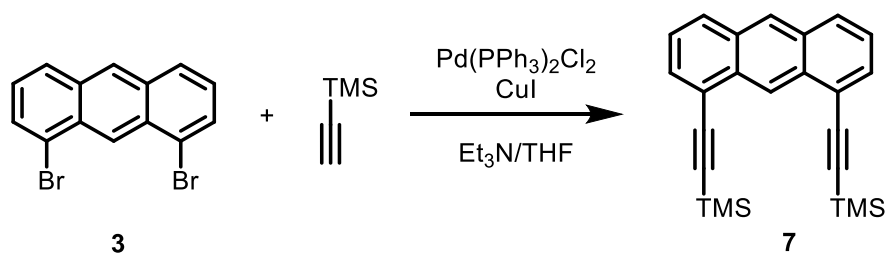

Figure S1. Synthesis of **7**

Dibromanthracene **3** (250 mg, 1.0 mmol), Pd(PPh)$_3$)$_2$Cl$_2$ (53 mg, 0.075 mmol) and CuI (8 mg, 0.075 mmol) were dissolved in a mixture of Et$_3$N/THF (1:1, 15 mL). Trimethylsilylacetylene (260 μL, 2.5 mmol) was added and the reaction mixture was stirred at 90 ºC for 24 h. Then, the solvent was evaporated under reduced pressure and the residue purified by column chromatography (SiO$_2$; hexane/CH$_2$Cl$_2$, 10:1), to isolate compound **7** (221 mg, 80%) as a yellowish solid. [1]

**¹H NMR** (300 MHz, CDCl₃) δ: 9.33 (s, 1H), 8.40 (s, 1H), 7.97 (d, $J$ = 8.7 Hz, 2H), 7.79 (d, $J$ = 6.3 Hz, 2H), 7.41 (dd, $J$ = 8.7, 6.6 Hz, 2H), 0.40 (s, 18H) ppm.

*Preparation of 1,8-diethynylanthracene (8)*

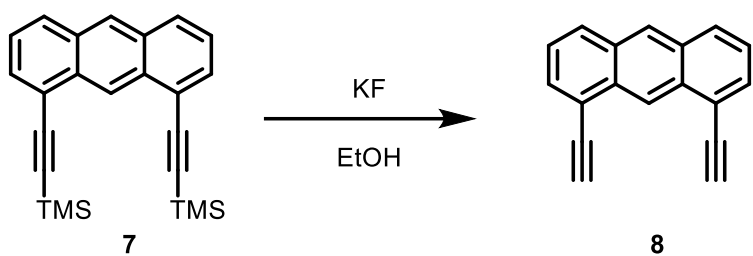

Figure S2. Synthesis of **8**

Compound **7** (275 mg, 0.74 mmol) and KF (215 mg, 3.74 mmol) were dissolved in ethanol (18 mL) and the reaction mixture was reflux for 2 h. Then, the solvent was evaporated under reduced pressure and the residue purified by column chromatography (SiO₂; hexane), to isolate diyne **8** (135 mg, 80%) as a white solid. [2]

**¹H NMR** (300 MHz, CDCl₃) δ: 9.43 (s, 1H), 8.41 (s, 1H), 7.99 (d, $J$ = 8.4 Hz, 2H), 7.79 (d, $J$ = 6.9 Hz, 2H), 7.43 (dd, $J$ = 8.4 Hz y 6.9 Hz, 2H), 3.63 (s, 2H) ppm.

*Preparation of ((8-ethynylanthracen-1-yl)ethynyl)triisopropylsilane (4)*

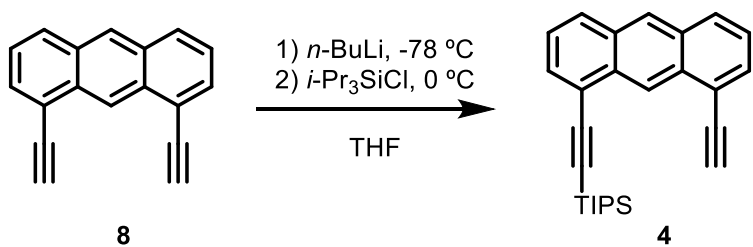

Figure S3. Synthesis of **4**

*n*-BuLi (196 µL, 0.49 mmol, 2.5 M) was added dropwise to a solution of diyne **8** (100 mg, 0.44 mmol) in THF (5 mL) at -78 °C and the mixture was stirred for 1.5 h at this temperature. Then, *i*-Pr₃SiCl (95 µL, 0.44 mmol) was added dropwise at 0 °C, the mixture was stirred for 3 h at 0 °C and allowed to warm up to room temperature for 16 h. Then H₂O was added, the organic phase was separated and the aqueous phase was extracted with CH₂Cl₂ (3 x 3 mL). The combined organic phases were dried over Mg₂SO₄. The solvent was evaporated under reduced pressure and the residue was purified by column chromatography (SiO₂; hexane), to afford **4** (140 mg, 83 %) as a yellow oil. [2]

**¹H NMR** (300 MHz, CDCl₃) δ: 9.64 (s, 1H), 8.45 (s, 1H), 8.03 (d, $J$ = 8.6 Hz, 1H), 8.00 (d, $J$ = 8.6 Hz, 1H), 7.79 (d, $J$ = 7.1 Hz, 2H), 7.44 (dd, $J$ = 7.1, 8.6 Hz, 2H), 3.49 (s. 1H), 1.16-1.24 (21H, m) ppm.

*Preparation of 1,4-bis(8-((triisopropylsilyl)ethynyl)anthracen-1-yl)buta-1,3-diyne (9)*

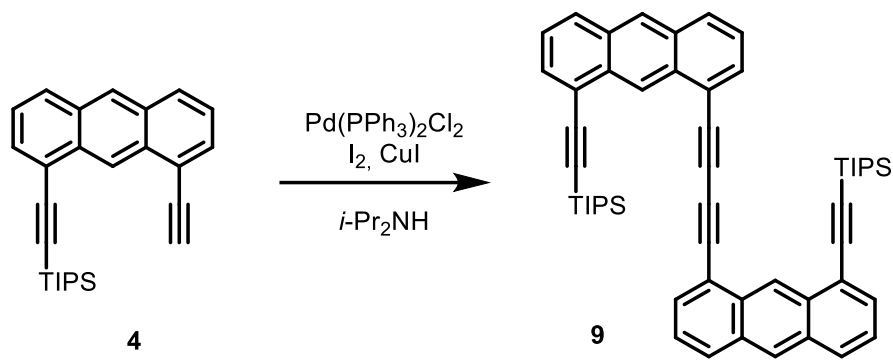

Figure S4. Synthesis of **9**

Compound **4** (230 mg, 0.60 mmol) was disolved in *i*-Pr₂NH (6 mL). Then, Pd(PPh₃)₂Cl₂ (21.0 mg, 30.0 μmol), CuI (12.0 mg, 60.0 μmol) and I₂ (76 mg, 30 μmol) were added and the reaction mixture was stirred at room temperature for 16 h. The solvent was evaporated under reduced pressure and the residue purified by column chromatography (SiO₂, hexane/CH₃Cl, 6:1) to afford compound **9** (80 mg, 35 %) as a yellow solid. [3]

**¹H NMR** (300 MHz, CDCl₃) δ: 9.50 (s, 2H), 8.46 (s, 2H), 8.05 (d, $J$ = 8.3 Hz, 2H), 7.99 (d, $J$ = 8.3 Hz, 2H), 7.89 (d, $J$ = 6.3 Hz, 2H), 7.78 (d, $J$ = 6.3 Hz, 2H), 7.48 (dd, $J$ = 6.8, 8.3 Hz, 2H), 7.44 (dd, $J$ = 6.8, 8.3 Hz, 2H), 1.12 (septet, $J$ = 6.3 Hz, 6H), 1.06 (d, $J$ = 6:3 Hz, 36H) ppm.

*Preparation of 1,4-bis(8-ethynylanthracen-1-yl)buta-1,3-diyne (1)*

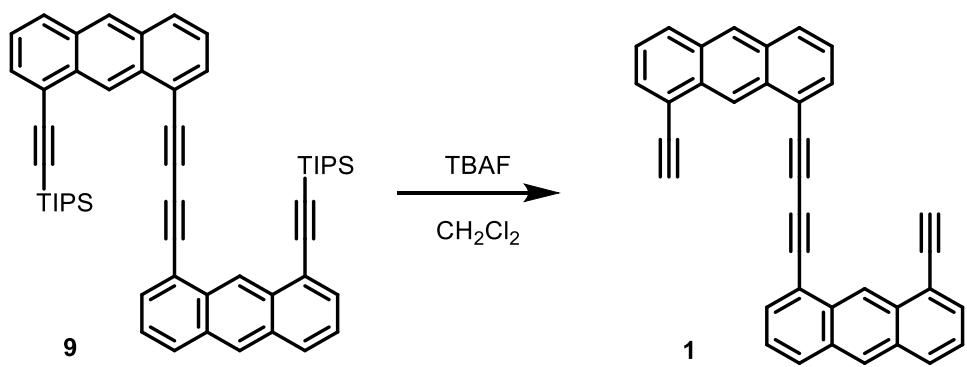

Figure S5. Synthesis of **1**

Over a solution of **9** (70 mg, 0.090 mmol) in CH₂Cl₂ (18 mL), TBAF (185 μL, 0.185 mmol, 1.0 M in THF) was added dropwise and the mixture was stirred at room temperature for 16 h.

Then, H₂O (3 mL) was added, the organic phase was separated and the aqueous phase was extracted with CH$_2$Cl$_2$ (3 x 5 mL). The combined organic phases were dried over Mg$_2$SO$_4$. The solvent was evaporated under reduced pressure and the residue was purified by column chromatography (SiO$_2$, hexane/CH$_3$Cl, 7:1), to isolate compound **1** (25 mg, 61 %) as a yellow solid.

**$^1$H NMR** (300 MHz, CDCl$_3$) δ: 9.53 (s, 2H), 8.49 (s, 2H), 8.06 (dd, $J$ = 8.7, 7.1 Hz, 4H), 7.92 (dd, $J$ = 7.0, 1.1 Hz, 2H), 7.81 (dd, $J$ = 6.9, 1.1 Hz, 2H), 7.53 − 7.44 (m, 4H), 3.70 (s, 2H) ppm.

**$^{13}$C NMR** (75 MHz, CDCl$_3$) δ: 132.62, 131.74, 131.57, 131.43, 131.31, 129.94, 129.41, 127.78, 125.13, 123.79, 120.46, 120.24, 83.44, 81.45, 79.57 ppm.

**MS (APCI (M+1)) HR** calculated for C$_{36}$H$_{19}$: 451.1481, found: 451.1482.

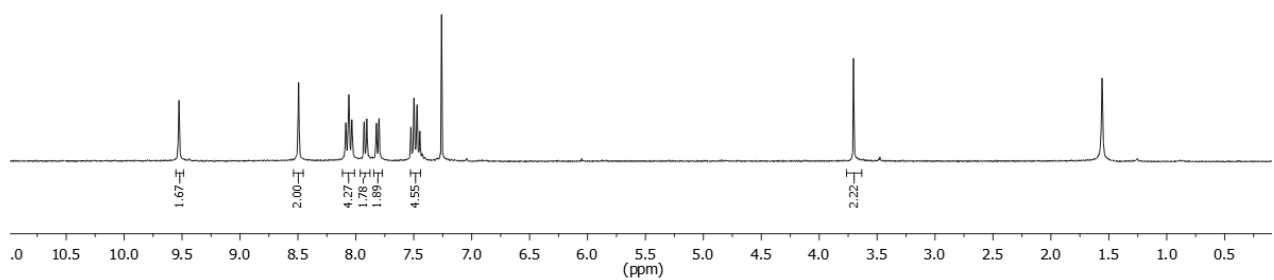

Figure S6. $^1$H NMR spectra for **1**

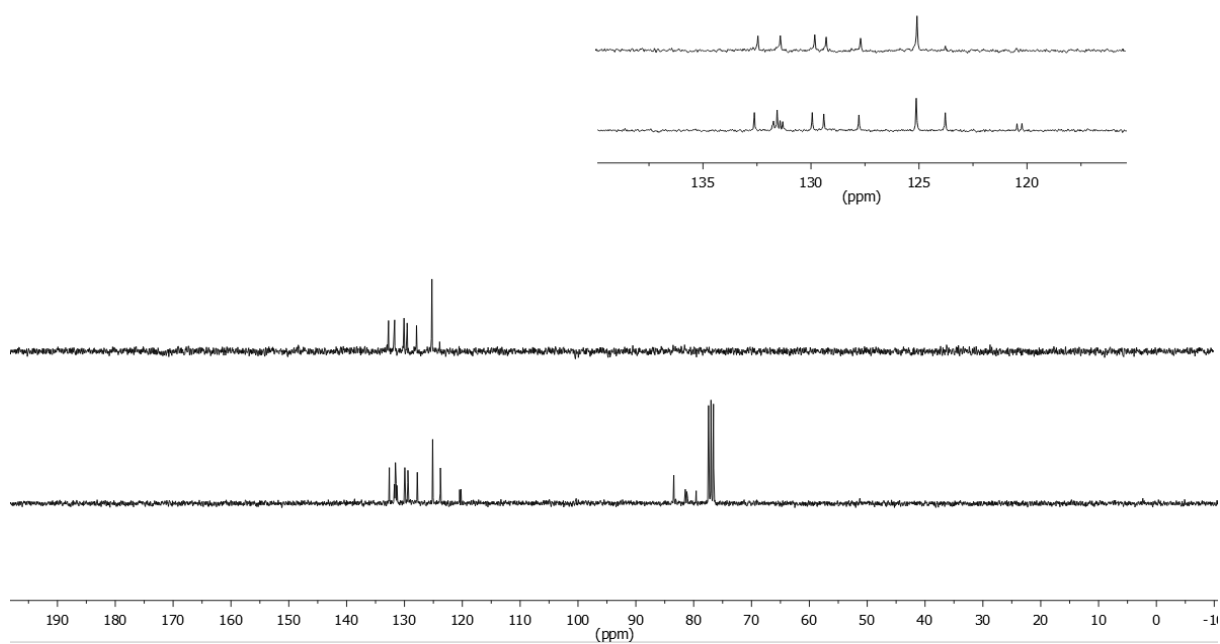

Figure S7. $^{13}$C NMR spectra for **1**

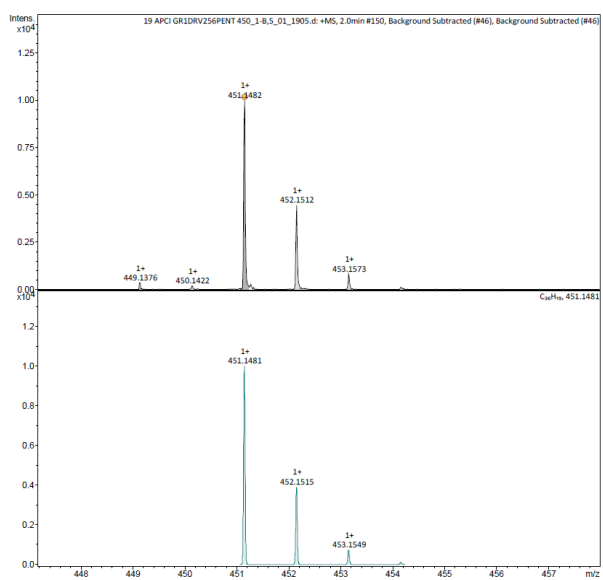

Figure S8. HRMS (APCI) for **1** (M+1).

**Details of DFT simulations:**

Density functional theory (DFT) simulations were performed using the FHI-AIMS code [4]. The structures of the reaction product as well as precursor *trans*-**1** were optimized in the gas phase. To mimic the molecule-substrate interaction effects onto the helical structure of the *cis*-**1** reactant, we relaxed this molecule on a bilayer NaCl slab of 8 x 8 atoms – consisting of 128 substrate atoms in total. The atom positions for the lower NaCl layer were constrained. All the simulations used the Perdew-Burke-Ernzerhof exchange-correlation functional [5].

**Additional AFM and STM measurements:**

Fig. S9 shows helicity switching of *cis*-**1** observed with the AFM.

Figure S10 shows a tip induced Glaser coupling reaction, where the molecule changed its adsorption site only a little, because it was adsorbed at a step edge to a third layer NaCl patch.

Figure S11 shows I/V and d$I$/d$V$ scanning tunneling spectroscopy above the product **2**.

Figure S12 and S13 show measured orbital densities of the precursors *cis*-**1** and *trans*-**1**, respectively. DFT calculated orbital densities are shown in addition.

Figure S13 shows the calculated LUMO, LUMO+1 and LUMO+2 of **2** and in comparison, those of diethynylanthracene.

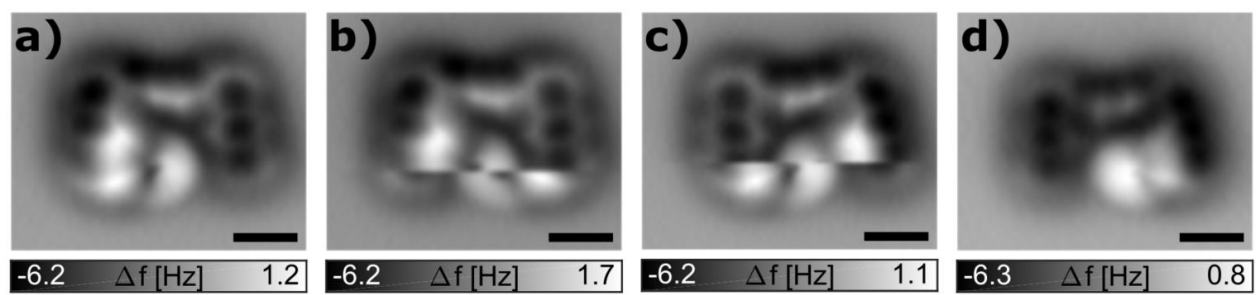

**Fig S9: Helicity switching of *cis*-1 while imaging.** a)-d) series of AFM images on the same individual molecule. a) and d) show the molecule in *(P)-cis-**1*** and *(M)-cis-**1*** conformation, respectively. b) and c) show AFM images during which the helicity was switched. The helicity of the molecule switched from *(P)-cis-**1*** to *(M)-cis-**1*** in the lower part of b). The molecule switched back from *(M)-cis-**1*** to *(P)-cis-**1*** in c). The slow scan direction was always from top to bottom. The images were recorded at tip-height offsets Δ$z$ = 0.9 Å, 0.8 Å, 0.9 Å and 1.0 Å for panels a), b), c) and d), respectively, with respect to the STM setpoint of $I$ = 1 pA at $V$ = 0.1 V on bare NaCl. All scale bars correspond to 5 Å.

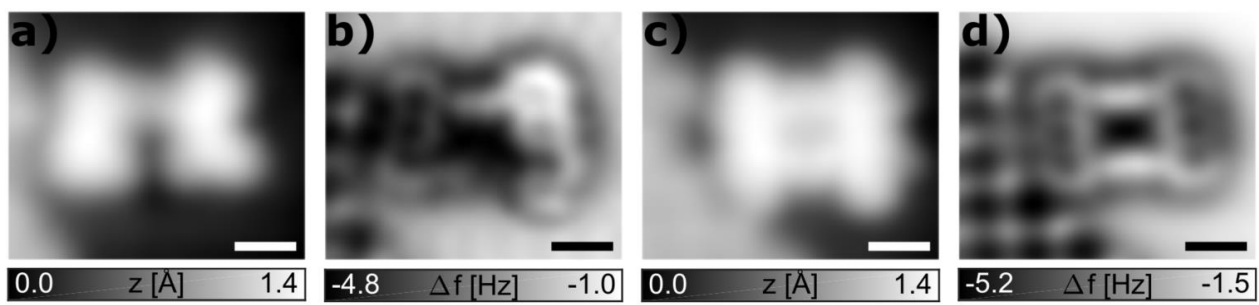

**Fig S10: Imaging of precursor *cis*-1 and product 2 adsorbed next to a patch of third layer NaCl, before and after the tip-induced coupling reaction.** a) shows a constant-current STM map and b) a constant-height AFM image (Δz = -0.10 Å) of precursor *cis*-**1**. c) and d) show STM and AFM measurements after applying a voltage pulse of 5.75 V, generating **2**. Parameters: a) and c) $I$ = 1 pA, $V$ = 1.25 V; b) and d) Δz = -0.10 Å and Δz = -0.15 Å, respectively, with respect to the STM setpoint of $I$ = 1 pA, $V$ = 1.25 V.

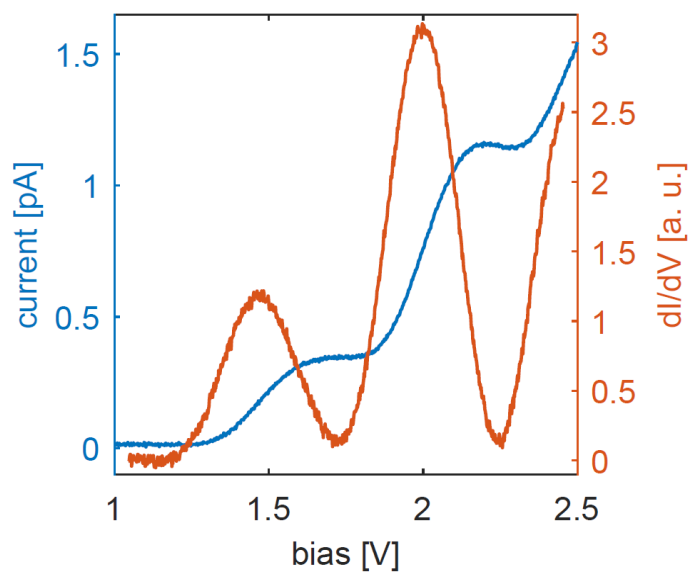

**Fig S11: Differential conductance spectrum of reaction product 2.** The blue curve shows the tunneling current $I$ recorded with a metal tip over the reaction product **2** as a function of sample bias voltage $V$. The red curve shows the numerically derived differential conductance curve (d$I$/d$V$). The peaks at 1.5 and 2.0 V are assigned to the LUMO and the LUMO+1 of the product **2**. The third feature, with onset at 2.3 $V$, is assigned to the LUMO+2.

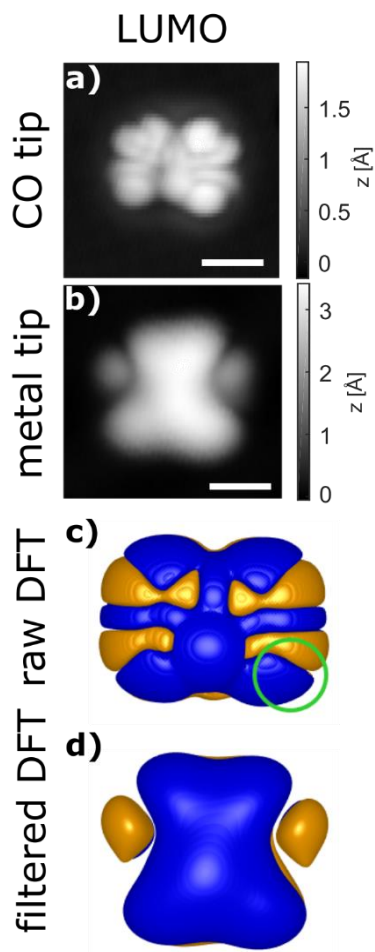

**Fig S12: Orbital density imaging and simulated orbitals of *(M)-cis*-1.** Constant-current STM images of the LUMO densities of *(M)-cis*-**1** adsorbed on bilayer NaCl with a) CO terminated tip and b) Cu terminated tip. Parameters: a) $I$ = 1 pA, $V$ = 1.5 V; b) $I$ = 1 pA, $V$ = 1.6 V. c) unprocessed DFT derived LUMO iso-surface and d) iso-surface of LUMO after applying a filter as indicated by the green circle in c). The molecule is oriented similar to Fig. 2d) of the main text.

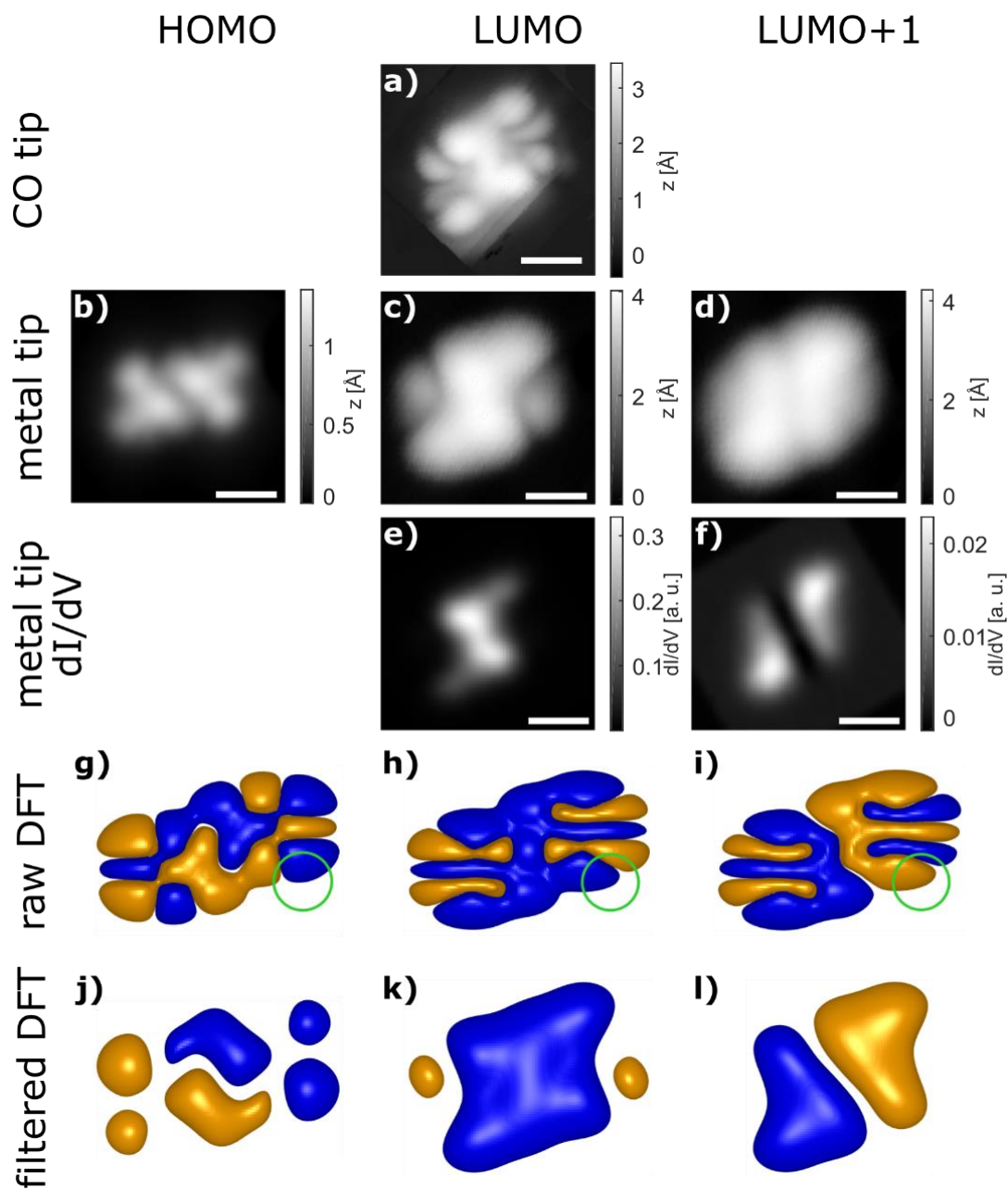

**Fig S13: Orbital density imaging and simulated orbitals of *trans*-1.** a)-f) show experimental orbital density imaging of the molecule adsorbed on bilayer NaCl with a) CO tip and b)-f) Cu tip. Parameters: a) $I$ = 1 pA, $V$ = 1.5 V; b) $I$ = 2 pA, $V$ = -2.8 V; c) $I$ = 1 pA, $V$ = 1.5 V; d) $I$ = 1 pA, $V$ = 2.1 V. e), f) d$I$/d$V$ images acquired at constant height with a sinusoidal modulation at 163 Hz and $V_{ac}$ = 50 mV applied in addition to the dc voltage $V_{dc}$ to extract the d$I$/d$V$ signal using lock-in technique. Parameters: e) Setpoint $I$ = 10 pA, $V$ = 1.5 V, $\Delta z$ = 2.5 Å, $V_{dc}$ = 1.5; f) Setpoint $I$ = 10 pA, $V$ = 2.0 V, $\Delta z$ = 4.0 Å, $V_{dc}$ = 2.0 V. g) – i) unprocessed DFT derived iso-surfaces of orbitals and j) – l) iso-surfaces of orbitals after applying a filter as indicated by the green circle in g)-i). The molecule is oriented similarly to Fig. 2b) of the main text.
18

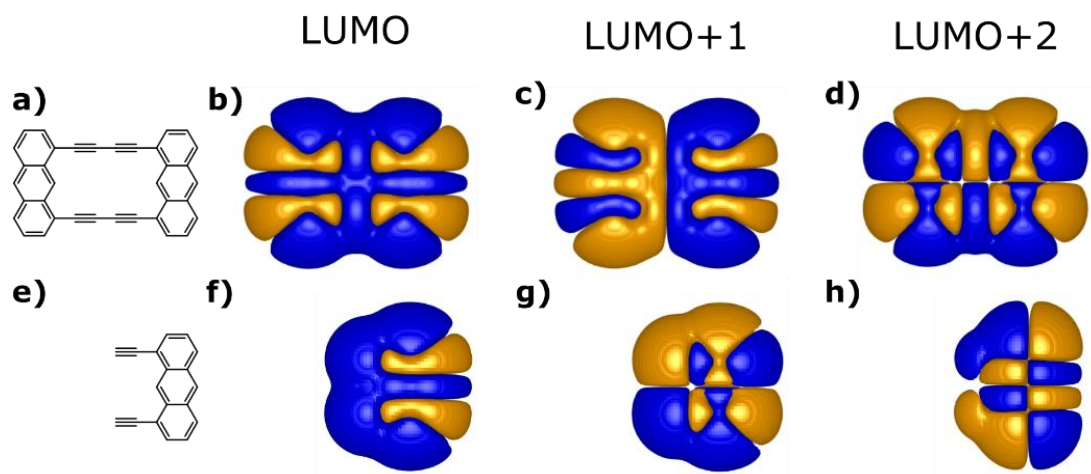

**Fig S14: DFT derived orbitals of 2 and diethynylanthracene.** b)-d) show the three lowest unoccupied orbitals of **2** and f)-h) those of diethynylanthracene. The models a) and e) provide the orientation of both molecules. Combining two LUMOs of diethynylanthracene in a bonding (symmetric) and antibonding (antisymmetric) manner rationalize the LUMO and LUMO+1 of **2**, respectively. The bonding combination of two LUMO+1 orbitals of diethynylanthracene rationalizes the LUMO+2 of **2**.